\documentclass[submission,copyright,creativecommons]{eptcs}
\providecommand{\event}{EXPRESS/SOS 2022} 
\title{
Bisimulations Respecting Duration and Causality for the Non-interleaving Applied $\pi$-Calculus
}
\def\titlerunning{Designing Semantics For the Non-interleaving Applied $\pi$-Calculus}

\author{%
	Cl{\'e}ment Aubert
	\institute{Augusta University, USA}
	\email{caubert@augusta.edu}
\and
	Ross Horne
	\institute{University of Luxembourg, Luxembourg}
	\email{ross.horne@uni.lu}
\and 
	Christian Johansen
	\institute{NTNU, Norway}
	\email{christian.johansen@ntnu.no}
}

\def\authorrunning{C. Aubert, R. Horne and C. Johansen}

\usepackage{underscore}         
\usepackage[T1]{fontenc}        
\usepackage{amsthm}
\newtheorem{definition}{Definition}

\usepackage{verbatim}
\usepackage{tikz}
\usetikzlibrary{positioning, calc}
\usepackage{csquotes}
\usepackage{xspace}
\usepackage{soul}
\usepackage{multicol}
\usepackage{mathtools}
\usepackage[export]{adjustbox}
\usepackage{lscape}
\usepackage{amsmath,amssymb}
\usepackage{tabularx}
\usepackage{array}
\usepackage[all]{xy}
\usepackage{latex/spi}
\usepackage{latex/prooftree}
\usepackage{breakurl}

\AtBeginDocument{

}

\newcommand*{\eg}{e.g.\@,\xspace}

\newcommand*{\cf}{cf.\@\xspace}
\newcommand*{\ie}{i.e.\@,\xspace}

\renewcommand{\st}{s.t.\@\xspace}
\newcommand*{\resp}{resp.\@\xspace}

\makeatletter
\newcommand*{\etc}{%
	\@ifnextchar{.}%
	{etc}%
	{etc.\@\xspace}%
}
\makeatother

\newcommand{\dlts}[2]{\updownlts^{#1}_{#2}}
\renewcommand{\dlts}[2]{\xrightarrow[\raisebox{.15em}{\protect{\(\scriptstyle #2\)}}]{#1}} 
\newcommand{\stld}[2]{\xleftarrow[\raisebox{.15em}{\protect{\(\scriptstyle #2\)}}]{#1}} 
\renewcommand{\lts}[1]{\xrightarrow{#1}}
\renewcommand{\pair}[2]{\left\langle#1, #2\right\rangle} 

\newcommand{\simi}[1]{\mathrel{\preceq_{#1}}}
\newcommand{\nsimi}[1]{\mathrel{\npreceq_{#1}}}
\newcommand{\presimi}[1]{\mathrel{\sqsubseteq_{#1}}}

\newcommand{\bisimi}[1]{\mathrel{\sim_{#1}}}
\newcommand{\nbisimi}[1]{\mathrel{\not\sim_{#1}}}
\newcommand{\fsimi}[1]{\mathrel{\preceq_{#1\kern-.1emf}}}
\newcommand{\nfsimi}[1]{\mathrel{\npreceq_{#1\kern-.1emf}}}

\newcommand{\alia}{\alpha}
\newcommand{\alib}{\beta}
\newcommand{\alic}{\gamma}

\newcommand{\match}[1]{\mathopen{\left[ #1 \right]}}

\newcommand{\prefix}[1]{\textcolor{red}{#1}}
\renewcommand{\n}[1]{\fv{#1}}

\newcommand{\Locnoarg}{\mathcal{L}\mkern-4mu oc}
\newcommand{\Loc}[1]{\mathopen{\Locnoarg}\left( #1 \right)}

\renewcommand{\names}[2]{#2} 
\renewcommand{\env}{\vec{\alic}} 

\newcommand{\fa}[1]{\mathopen{\mathrm{fa}}\left(#1\right)} 
\newcommand{\aliroot}{\lambda} 

\DeclareMathOperator{\id}{id} 
\DeclareMathOperator{\pok}{ok}
\DeclareMathOperator{\perr}{er}
\DeclareMathOperator{\prhi}{hi}

\newcommand{\Indy}{\smile} 
\newcommand{\notIndy}{\mathrel{\not\smile}} 
\newcommand{\sIndy}{\mathrel{I_{\ell}}}
\newcommand{\ESS}{\mathrel{\mathsf{S}}}
\newcommand{\ESSTEE}{\mathcal{E}}

\definecolor{green(ncs)}{rgb}{0.0, 0.62, 0.42}
\newcommand{\diff}[1]{\textcolor{green(ncs)}{#1}} 	

\newcommand\tworels[2]{\genfrac{}{}{0pt}{}{#1}{#2}}

\newcommand{\sizeof}[1]{\left| #1 \right|}

\definecolor{clem}{HTML}{60656F} 
\definecolor{ross}{HTML}{BABD8D} 
\definecolor{chri}{HTML}{EB6424} 
\newcommand{\clem}[1]{\textcolor{clem}{#1}}
\newcommand{\ross}[1]{\textcolor{ross}{#1}}
\newcommand{\chri}[1]{\textcolor{chri}{#1}}

\begin{document}
\maketitle
  
\begin{abstract}
This paper shows how we can make use of an asynchronous transition system, whose transitions are labelled with events and which is equipped with a notion of independence of events,
to define non-interleaving semantics for the applied $\pi$-calculus.
The most important notions we define are: Start-Termination or ST-bisimilarity, preserving duration of events;
and History-Preserving or HP-bisimilarity, preserving causality.
We point out that corresponding similarity preorders expose clearly 
distinctions between these semantics.
We draw particular attention to the distinguishing power of HP failure similarity,
and discuss how it affects the attacker threat model
against which we verify security and privacy properties.
We also compare existing notions of 
located bisimilarity to the definitions we introduce.
\end{abstract}

\section{Introduction}


Non-interleaving semantics is sometimes referred to as true concurrency.
This reflects the idea that parallel composition has a semantically distinct
status from its interleavings obtained by allowing each parallel process
to preform actions one-by-one in any order. 
In this work, we explore a spectrum of non-interleaving semantics for the applied $\pi$-calculus,
which is
motivated by 
some recent works on 
modelling and verifying security and privacy properties of cryptographic protocols~\cite{Cheval2019,Horne2021}.
The definitions introduced are operational in style, bypassing denotations such as event structures.

We build on our recent work~\cite{Aubert2022e} that introduced 
a non-interleaving Structural Operational Semantics (SOS) for the applied $\pi$-calculus that 
generates 
Labelled Asynchronous Transition Systems (LATS).
Compared with standard transition systems, whose transitions are labelled with actions,
a LATS labels its transitions with richer \emph{events},
and is equipped with a notion of \emph{independence} over adjacent events (concurrently enabled or enabled one after another). 
A LATS allows independent events to be permuted and hence techniques such as partial-order reduction to be applied.
This work is part of a research agenda where
we wish to 
lay a foundation for exploring questions such as whether
verification techniques are enabled by adopting a semantics
that is naturally compatible with an independence relation 
used for partial-order reduction.
Another research question is 
whether adopting a
non-interleaving semantics impacts the attacker model for certain
problems.
In particular, armed with our definitions, we may ask whether our non-interleaving semantics
may detect attacks that may be missed if we employ an interleaving semantics.

The contribution of this paper towards addressing the questions above is the introduction of non-interleaving equivalences and similarities that can be defined for the applied $\pi$-calculus
equipped with a LATS~\cite{Aubert2022e}.
A well understood starting point is how to generate 
\enquote{located} equivalences~\cite{Boudol1994} 
for CCS~\cite{Castellani1995, Mukund1992} and the $\pi$-calculus~\cite{Sangiorgi1996}.
The former approach makes direct use of the LATS for CCS, while the latter uses a cut down located transition system for the $\pi$-calculus which accounts for locations but does not satisfy all properties of a LATS.
We go further since, given our LATS,
we can
generate in an operational style other notions of non-interleaving semantics, particularly
those that preserve duration of events (Start-Termination or ST semantics)~\cite{Glabbeek90TRb} and those that preserve causality (History-Preserving or HP semantics)~\cite{Degano1988b,Rabinovich1988}.
Since we cover the applied $\pi$-calculus, of course, we encompass the $\pi$-calculus,
where the later surprisingly benefits from adopting a modern applied $\pi$-calculus style when handling \emph{link causality}
-- the causal relationship between outputs and inputs that depend upon them.
Our operational approach also avoids the need to unfold to event structures~\cite{Crafa2012,Varacca2010} or
configuration structures~\cite{Cristescu2015b} that would track entire histories of causal dependencies;
instead, we consider only what is happening or enabled at a particular point in time.

We include in Tables \ref{tb1} and \ref{tb2} a glossary, including key standard
and non-standard terminology employed in this paper.
We emphasise similarity rather than bisimilarity for two reasons.
Firstly, similarity exposes more clearly than bisimilarity the differences between
non-interleaving semantics as it allows clearer separating examples.
Secondly, similarity is known to have compelling attacker models in terms
of probabilistic may testing~\cite{Deng2008}, and it is standard in computational security
to consider probabilistic attackers~\cite{Cheval2022}.
\autoref{tb1} presents the notions of 
similarity that we discuss in the interleaving/non-interleaving
spectrum we explore.
Along this spectrum the attacker has different powers for observing 
concurrency. 

\begin{table}
\begin{tabularx}{\linewidth-1pt}{|l|X|l|} 
\hline
\emph{Terminology} & \emph{Remarks} & \emph{Def.} 
\\\hline
 i-similarity        & \enquote{Interleaving}-similarity is the notion of similarity most commonly explored in the literature. & \mbox{\autoref{def:i-sim}}
\\\hline
 ST-similarity        & \enquote{Start-Terminate}-similarity accounts for the fact that events have duration. It uses events to distinguish between actions with the same label, and to ensure that two \enquote{terminate} events correspond to the same \enquote{start} event. & \mbox{\autoref{def:st-sim}}
\\\hline
 HP-similarity        & \enquote{History-Preserving}-similarity preserves the causal dependencies between events. & \mbox{\autoref{def:hp-sim}}
\\\hline
 $I$-similarity  & \enquote{Independence}-similarity are parametrised by some notion of independence $I$. We obtain \enquote{located bisimilarities} using the structural independence relation $\sIndy$ that considers only if two events are in different locations. 
  & \mbox{\autoref{def:l-sim}}
\\ \hline
\end{tabularx}
\caption{Strategies in the interleaving/non-interleaving spectrum explored for the applied $\pi$-calculus.}
\label{tb1}
\end{table}

While we draw attention to similarity, 
we are also interested in non-interleaving \emph{bi}similarity and other notions in the linear-time/branching-time spectrum~\cite{glabbeek90concur}.
Indeed, all the notions in \autoref{tb1} also exist in their other variants in the linear-time/branching time spectrum listed in \autoref{tb2}, such as failure similarity.
Along this spectrum the observer has more or less power to observe and make choices.
We also use the term \emph{mutual}, \eg mutual ST-similarity, when some notion of similarity holds in both directions.

\begin{table}
\begin{tabularx}{\linewidth}{|l|X|c|} 
\hline
 \emph{Terminology} & \emph{Remarks} & \emph{Symb.} 
 \\ \hline
 X-bisimilarity      & An equivalence ranging over all strategies of a particular type X. & \(\bisimi{X}\)
 \\ \hline
 X-similarity        & The preorder arising when we assume one player leads throughout a strategy (except when testing equations, as explained around \autoref{def:i-sim}). & \(\simi{X}\)
 \\ \hline
 X-presimilarity     & A notion of similarity we introduce in this paper (\autoref{def:presim}) to emphasise the testing power of inequalities in the applied $\pi$-calculus. & \(\presimi{X}\)
 \\ \hline
 Xf-similarity       & X \enquote{failure} similarity is one of many variants of similarity in the linear-time/branching-time spectrum, and is chosen due to its testing model allowing us to test if something is not enabled. In particular, we look at STf-similarity (\autoref{def:STf-sim}) and HPf-similarity (\autoref{def:HPf-sim}).& \(\fsimi{X}\)
\\\hline
\end{tabularx}
\caption{Notions in the linear-time/branching-time spectrum explored for the applied $\pi$-calculus.}
\label{tb2}
\end{table}

There are further spectra that could be explored: for the $\pi$-calculus there is the open/early spectrum, including notions such as early, late, quasi-open~\cite{Sangiorgi2001c}, and open~\cite{Sangiorgi1996c} variants of equivalences.
This work considers only \emph{early} and  \emph{strong} semantics: early semantics means that the message input is chosen at the moment the event starts,
whereas the other variants allow different degrees of laziness in learning what message was input retrospectively.
This choice is made since the majority of equivalences for the applied $\pi$-calculus in the literature are early, 
and early bisimilarity coincides with notions of testing via concurrent processes~\cite{Abadi2018}.
Since our semantics are strong, every $\tau$-transition
is matched by exactly one $\tau$-transition in all our strategies.
Many security and privacy problems
that motivate us can be reduced to a
strong equivalence problem.
However, the main reason for these choices is simply to focus on the interleaving/non-interleaving spectrum.
For example, it would be easy to define quasi-open variants of our non-interleaving semantics, which
coincide with a testing semantics making use of all contexts~\cite{Yurkov2021}.

 After briefly recalling our non-interleaving SOS generating a LATS (\autoref{sec:SOS}), we use interleaving semantics to illustrate and motivate the genericity of \emph{static} equivalences (\autoref{sec:interleaving}).
\autoref{sec:semantics-lats} is the core of our proposal: it starts by introducing and stressing the importance of the independence relation (\autoref{ssec:inde}), which is used throughout the rest of the article.
ST and HP-similarities are then defined in Sect.~\ref{ssec:st} and \ref{ssec:hp}
and compared in the context of privacy in \autoref{ssec:privacy}.
\autoref{ssec:failure} discusses failure semantics for HP- and ST-similarities.
Some design decisions are justified in light of located bisimulations in \autoref{sec:located-bisim}.

\section{Background: A Non-interleaving SOS for the Applied \texorpdfstring{\(\pi\)}{pi}-Calculus}\label{sec:SOS}

This section recalls a non-interleaving structural operational semantics for the applied $\pi$-calculus.
The design decisions are discussed extensively in a companion paper~\cite{Aubert2022e}.
What we present below is intended only as a condensed summary of that operational semantics for ease of reference.

All variables $x, y, z$ are the same syntactic category,
but are distinct from \emph{aliases}.
Aliases range over $\alia, \alib, \alic$ and
consist of an alias variable, say $\lambda$,
prefixed with a string $\prefix{s} \in \left\{ \prefix{0}, \prefix{1} \right\}^*$,
\ie $\alia = \prefix{s}\lambda$.
\emph{Messages} range over $M, N, K$, built from a signature of function symbols $\Sigma$.
As standard, a \emph{substitution} $\sigma, \theta$ or $\rho$ is a function with a 
domain ($\dom{\sigma} = \left\{ \alia \colon \alia \neq \alia\sigma \right\}$) and a range ($\ran{\sigma} = \left\{\alia\sigma \colon \alia \in \dom{\sigma} \right\}$) 
that are applied in suffix form.
The \emph{identity substitution} is denoted \(\id\)
and
composition $\sigma \circ \theta$.

\begin{figure}
	\begin{multicols}{2}
		$\begin{array}{rlr}
			\multicolumn{3}{l}{\textsc{Processes:}}\\
			P, Q, R \Coloneqq& 0 & \mbox{deadlock} \\
			\mid& \mathopen\nu x. P & \mbox{new} \\
			\mid& P \cpar Q & \mbox{parallel} \\
			\mid& G & \mbox{guarded process} \\
			\mid& \bang P & \mbox{replication} 
			\\[1em]
			\multicolumn{3}{l}{\textsc{Guarded processes:}}\\
			G, H \Coloneqq& \cin{M}{x}.P & \mbox{input prefix} \\
			\mid &  \cout{M}{N}.P & \mbox{output prefix} \\
			\mid& \mathopen{\left[M = N\right]}G & \mbox{match} \\
			\mid& \mathopen{\left[M \neq N\right]}G & \mbox{mismatch} \\
			\mid& G + H & \mbox{choice} 
		\end{array}$
	
	$\begin{array}{rlr}
		\multicolumn{3}{l}{\textsc{Extended processes:}}\\
		A, B  \Coloneqq& \sigma \cpar P &   \mbox{active process} \\ 
		\mid& \mathopen{\nu x.} A &  \mbox{new}
	\\[1em]
		\multicolumn{3}{l}{\textsc{Messages:}}\\
		M, N \Coloneqq& x & \mbox{variable} \\
		\mid &  \alia & \mbox{alias} \\
		\mid & f(M_1, \hdots, M_n) & \mbox{function} 
	\\[1em]
		\multicolumn{3}{l}{\textsc{Early action labels:}}\\
		\pi  \Coloneqq& M\,N &   \mbox{free input} \\ 
		\mid& \co{M}(\alia) &  \mbox{output} \\
		\mid& \tau &  \mbox{interaction} 
		\end{array}$
	\end{multicols}
	\caption{Syntax of extended processes with guarded choices, where $f \in \Sigma$.            
	}\label{figure:syntax}
\end{figure}
%

Processes are denoted by $P, Q, R$, and in $\nu x.P$ and $a(x).P$ occurrences of $x$ in $P$ are bound.
Sequences of names $\nu \vec{x}. P$ abbreviate multiple name binders defined inductively such that $\nu \epsilon. P = P$ and $\nu x, \vec{y}. P = \nu x. \nu \vec{y}.P$, where $\epsilon$ is the empty sequence.
\emph{Active substitutions}, denoted $\sigma$, $\theta$, map aliases in their finite domain to messages containing no aliases,
and appear in \emph{extended processes}, ranging over $A, B, C$.
We assume a \emph{normal form}, where aliases do not appear in processes, and an \emph{equational theory} $E$ containing equalities on messages, \eg $\dec{\enc{M}{K}}{K} \mathrel{=_E} M$.
Figs.\ \ref{figure:syntax} and \ref{figure:active} give the syntax and semantics.

\begin{definition}[freshness, $\alpha$-equivalence, etc.]
	A variable \(x\) (\resp an alias \(\alia\)) is \emph{free in a message \(M\)} if \(x \in \fv{M}\) (\resp \(\alia \in \fa{M}\)) for
	\begin{align*}
	\fv{f(M_1, \ldots M_n)} &= \cup_{i = 1}^n \fv{M_i}
	&&&
	\fv{x} &= \left\{ x \right\}
	&&&
	\fv{\alia} &= \emptyset
	\\
	\fa{f(M_1, \ldots M_n)} &= \cup_{i = 1}^n \fa{M_i}
	&&& 
	\fa{x} &= \emptyset
	&&&
	\fa{\alia} &= \left\{ \alia \right\}.
	\end{align*}
	The \(\mathrm{fv}\) function extends in the standard way to (extended) processes, letting $\fv{\nu x.P} = \fv{P}\setminus\left\{x\right\}$
	and $\fv{M(x).P} = \fv{M} \cup \left(\fv{P}\setminus\left\{x\right\}\right)$, and similarly for \(\fv{A}\).
The functions for free variables and free aliases extend to labels as follows.
\[
\begin{gathered}
\begin{array}{c}
\fv{\pi}
=
\left\{ 
\begin{array}{ll}
\fv{M}\cup\fv{N} & \mbox{if $\pi = M\,N$}
\\
\fv{M}           & \mbox{if $\pi = \co{M}(\alia)$}
\\
\emptyset        & \mbox{if $\pi = \tau$}
\end{array}
\right.
\hspace{2em} \fa{\pi}
=
\left\{ 
\begin{array}{ll}
\fa{M}\cup\fa{N} & \mbox{if $\pi = M\,N$}
\\
\fa{M}           & \mbox{if $\pi = \co{M}(\alia)$}
\\
\emptyset        & \mbox{if $\pi = \tau$}
\end{array}
\right.
\end{array}
\end{gathered}
\]

	We say a variable $x$ is \emph{fresh for a message \(M\) (\resp process \(P\), extended process \(A\))}, written \(\isfresh{x}{M}\) (\resp \(\isfresh{x}{P}\), \(\isfresh{x}{A}\)) 	whenever $x \notin \fv{M}$ (\resp \(x \notin \fv{P}\), \(x \notin \fv{A}\)), and similarly for aliases.
	Freshness extends point-wise to lists of entities, \ie $\isfresh{x_1,x_2, \ldots x_m}{M_1, M_2, \ldots, M_n}$, denotes the conjunction of all $\isfresh{x_i}{M_j}$ for all $1 \leq i \leq m$ and $1 \leq j \leq n$.

	We define \emph{$\alpha$-equivalence} (denoted $\equiv_\alpha$) for variables only (not aliases which are fixed \enquote{addresses}) as the least congruence (a reflexive, transitive, and symmetric relation preserved in all contexts) such that,
	whenever $\isfresh{z}{\nu x.P}$, we have
	$\nu x.P \mathrel{\equiv_\alpha} \nu z.(P\sub{x}{z})$
	and $M(x).P \mathrel{\equiv_\alpha} M(z).(P\sub{x}{z})$.
	Similarly, for extended processes, we have the least congruence such that, whenever $\isfresh{z}{\nu x.A}$, we have
	$\nu x.A \mathrel{\equiv_\alpha} \nu z.(A\sub{x}{z})$.
	Restriction is such that $\theta\mathclose{\restriction_{\vec{\alpha}}}(x) = \theta(x)$ if $x \in \vec{\alpha}$ and $x$ otherwise.

	\emph{Capture-avoiding substitutions} are defined for processes such that $(M(x).P)\sigma \mathrel{\equiv_\alpha} M\sigma(z).P\sub{x}{z}\sigma$ and $(\nu x.P)\sigma \mathrel{\equiv_\alpha} \nu z.P\sub{x}{z}\sigma$ for $\isfresh{z}{\dom{\sigma}, \ran{\sigma}, \nu x.P}$.
	For extended processes, it is defined such that $(\nu x.A)\mathclose{\rho} \mathrel{\equiv_\alpha} \nu z.(A\mathclose{(\sub{x}{z} \circ \rho)})$
	and $(\sigma \cpar P)\rho = ({\sigma \circ \rho}\mathclose{\restriction_{\dom{\sigma}}} \cpar P\rho)$,
	for $\isfresh{z}{\dom{\rho}, \ran{\rho}, \nu x.A}$.
\end{definition}

\begin{definition}[structural congruence]\label{def:struct}
	Our minimal \emph{structural congruence} (denoted $\equiv$) is the least equivalence relation on extended processes 
	extending 
	$\alpha$-equivalence
	such that whenever $\sigma = \theta$, $P \mathrel{\equiv_\alpha} Q$ and $A \equiv B$, we have: 
	$\sigma \cpar P \equiv \theta \cpar Q$,
	$\nu x.A \equiv \nu x.B$
	and
	$\nu x.\nu z.A \equiv \nu z.\nu x.A$.
\end{definition}

\begin{figure*}[t]{
		\centering
		\setlength\extrarowheight{15pt}
		
		\begin{tabular}{c c}
			\begin{prooftree}
				M \mathrel{=_E} K
				\qquad
				\justifies
				\names{\env}{
					\mathopen{\cin{K}{x}.}P 
					\dlts{M\,N}{
						[]
					}
					{\id \cpar P\sub{x}{N}} }
				\using
				\mbox{\textsc{Inp}}
			\end{prooftree}
			&
			\begin{prooftree}
				M \mathrel{=_E} K
				\justifies
				\names{\env}{
					\cout{K}{N}.P 
					\dlts{\co{M}(\aliroot)}{
						[]
					}
					{\sub{\aliroot}{N}} \cpar P }
				\using
				\mbox{\textsc{Out}}
			\end{prooftree}
			\\
			\begin{prooftree}
				\names{\env}{
					P}
				\dlts{\pi}{u}
				\mathopen{\nu \vec{x}.}\left( \sigma \cpar R \right)
				\quad
				\isfresh{\vec{x}}{Q}
				\justifies
				\names{\env}{
					P \cpar Q
					\dlts{\pi}{\prefix{0}u}
					\mathopen{\nu \vec{x}.}\left( \sigma \cpar R \cpar Q \right)
				}
				\using
				\mbox{\textsc{Par-L}}
			\end{prooftree}
			&			\begin{prooftree}
				\names{\env}{
					Q}
				\dlts{\pi}{u}
				\mathopen{\nu \vec{x}.}\left( \sigma \cpar R \right)
				\quad
				\isfresh{\vec{x}}{P}
				\justifies
				\names{\env}{
					P \cpar Q
					\dlts{\pi}{\prefix{1}u}
					\mathopen{\nu \vec{x}.}\left( \sigma \cpar P \cpar R \right)
				}
				\using
				\mbox{\textsc{Par-R}}
			\end{prooftree}
			\\
			\begin{prooftree}
				\names{\env}{
					P\sub{x}{z}
					\dlts{\pi}{
						u 
					}
					A
				}
				\qquad
				\isfresh{z}{\n{\pi}, \nu{x}.{ P }}
				\justifies
				\names{\env}{
					\nu{x}.{ P }
					\dlts{\pi}{
						u 
					}
					\mathopen{\nu{z}.}A
				}
				\using
				\mbox{\textsc{Extrude}}
			\end{prooftree}
			&
			\begin{prooftree}
				\names{\env}{
					A
					\dlts{\pi}{
						u 
					}
					B
				}
				\qquad
				\isfresh{x}{\n{\pi} }
				\justifies
				\names{\env}{
					\nu{x}.{ A }
					\dlts{\pi}{
						u 
					}
					\mathopen{\nu{x}.}B
				}
				\using
				\mbox{\textsc{Res}}
			\end{prooftree}
			\\
			\multicolumn{2}{c}{
				\begin{prooftree}
					\names{\env}{
						P}
					\dlts{\co{M\sigma}(\lambda)}{\prefix{s}[s']} 
					\mathopen{\nu \vec{x}.}\left( {\sub{\lambda}{N}} \cpar Q \right)
					\quad
					\isfresh{\vec{x}}{\ran{\sigma}}
					\quad
					\fa{M} \subseteq \dom{\sigma}
					\quad
					\isfresh{\prefix{s}\lambda}{\dom{\sigma}}
					\justifies
					\names{\env}{
						\sigma \cpar P}
					\dlts{\co{M}(\prefix{s}\lambda)}{\prefix{s}[s']}
					\mathopen{\nu \vec{x}.}\left( \sigma \circ {\sub{\prefix{s}\lambda}{N}} \cpar Q \right)
					\using
					\textsc{Alias-out}
				\end{prooftree}
			}
			\\
			\multicolumn{2}{c}{
				\begin{prooftree}
					\names{\env}{
						P}
					\dlts{\pi\sigma}{u} 
					\mathopen{\nu \vec{x}.}\left( \id \cpar Q\right)
					\quad
					\isfresh{\vec{x}}{\ran{\sigma}}
					\quad
					\fa{\pi} \subseteq \dom{\sigma}
					\justifies
					\names{\env}{
						\sigma \cpar P}
					\dlts{\pi}{u}
					\mathopen{\nu \vec{x}.}\left( \sigma \cpar Q \right)
					\using
					\textsc{Alias-free}
				\end{prooftree}
			}
			\\
			\multicolumn{2}{c}{
			\setlength\extrarowheight{15pt}
			\begin{tabular}{c c c }
			\begin{prooftree}
				\names{\env}{ G } \dlts{\pi}{[t]} A
				\justifies
				\names{\env}{ G + H } \dlts{\pi}{[0t]} A
				\using
				\mbox{\textsc{Sum-L}}
			\end{prooftree}
			& \hfill
			\begin{prooftree}
				\names{\env}{ H } \dlts{\pi}{[t]} A
				\justifies
				\names{\env}{ G + H } \dlts{\pi}{[1t]} A
				\using
				\mbox{\textsc{Sum-R}}
			\end{prooftree}
\hfill
		&
			\begin{prooftree}
				\names{\env}{  P \cpar \bang P 
					\dlts{\pi}{u}
					A }
				\justifies
				\names{\env}{ \bang P 
					\dlts{\pi}{u}
					A }
				\using
				\mbox{\textsc{Bang}}
			\end{prooftree}
		\end{tabular}
	}
			\\
			\begin{prooftree}
				\names{\env}{ P } \dlts{\pi}{u} A
				\qquad
				M \mathrel{=_E} N
				\justifies
				\names{\env}{ {\mathopen{\left[M=N\right]}{P} }} \dlts{\pi}{u} 
				{A}
				\using
				\mbox{\textsc{Mat}}
			\end{prooftree}
			&
			\begin{prooftree}
				\names{\env}{ P } \dlts{\pi}{u} A
				\qquad\qquad
				M \mathrel{\not=_{E}} N 
				\justifies
				\names{\env}{ {\mathopen{\left[M \not= N\right]}{P} }}
				\dlts{\pi}{u}
				{A}
				\using
				\mbox{\textsc{Mismat}}
			\end{prooftree}
\\
			\multicolumn{2}{c}{
		\begin{prooftree}
			\names{\env}{ P }
			\dlts{\co{M}(\aliroot)}{\ell_0} 
			\mathopen{\nu \vec{y}.}\left( {\sub{\aliroot}{N}} \cpar P' \right)
			\qquad
			\names{\env}{  Q }
			\dlts{M\,N}{\ell_1}
			\mathopen{\nu \vec{w}.}\left(\id \cpar Q' \right)  
			\qquad
			\isfresh{\vec{y}}{Q}
			\qquad
			\isfresh{\vec{w}}{P, \vec{y}}
			\justifies
			\names{\env}{  P \cpar Q}
			\dlts{\tau}{(\prefix{0}\ell_0, \prefix{1}\ell_1)}
			\mathopen{\nu \vec{y}, \vec{w}.}\left(\id \cpar P' \cpar Q' \right) 
			\using
			\mbox{\textsc{Close-L}}
		\end{prooftree}
}
\\
			\multicolumn{2}{c}{
		\begin{prooftree}
			\names{\env}{ P }
			\dlts{M\,N}{\ell_0}
			\mathopen{\nu \vec{y}.} \left(\id \cpar P' \right)  
			\qquad
			\names{\env}{  Q }
			\dlts{\co{M}(\aliroot)}{\ell_1} 
			\mathopen{\nu \vec{w}.}\left(  {\sub{\aliroot}{N}} \cpar Q' \right)
			\qquad
			\isfresh{\vec{w}}{P}
			\qquad
			\isfresh{\vec{y}}{Q, \vec{w}}
			\justifies
			\names{\env}{  P \cpar Q}
			\dlts{\tau}{(\prefix{0}\ell_0, \prefix{1}\ell_1)}
			\mathopen{\nu \vec{y}, \vec{w}.}\left( \id \cpar P' \cpar Q' \right)
			\using
			\mbox{\textsc{Close-R}}
		\end{prooftree}
}
		\end{tabular}}
	\caption{
		An early non-interleaving structural operational semantics.
	}\label{figure:active}
\end{figure*}

\begin{definition}[location labels]\label{def:loclabel}
	A \emph{location} \(\ell\) is of the form $\prefix{s}[t]$, where $\prefix{s} \in \left\{\prefix{0},\prefix{1}\right\}^*$
	and $t \in \left\{{0},{1}\right\}^*$. 
	If $\prefix{s}$ or \(t\) is empty, we omit it (hence, we write \(\prefix{\epsilon}[\epsilon]\) as \([]\)).
	A \emph{location label} $u$ is either a location \(\ell\) or a pair of locations \((\ell_0, \ell_1)\), and we let $\prefix{c}(\ell_0, \ell_1) = (\prefix{c}\ell_0, \prefix{c}\ell_1)$ for $\prefix{c} \in \left\{\prefix{0},\prefix{1}\right\}$.
\end{definition}

\section{Handling located aliases, explained using interleaving similarities} \label{sec:interleaving}

Although the objective of this paper is to explore non-interleaving semantics,
we begin by defining an interleaving semantics.
The reason is that we wish to expose clearly which parts of our definitions are
generic to any type of semantics, and which are specific to non-interleaving
semantics.

The first shared trait by all equivalences for the applied $\pi$-calculus
is that they make use of a \emph{static equivalence}.
Its role is to prevent the attacker from using the data they know to form a test for one process that does not hold for another process.
In an extended process, one can think of the active substitution
as a record of the information available to an attacker observing messages communicated on public channels.
The attacker can then combine that information in various ways to try to pass a test, \eg hashing the first message and checking whether it is equal to the second message.
We find it insightful to break down static equivalence into simpler definitions,
that we will employ to achieve the same effect.
In particular, we start with the following satisfaction relation.
\begin{definition}[satisfaction]
Satisfaction $\vDash$ is defined inductively as:
\begin{itemize} 
\item
$\nu x.A \vDash M = N$ whenever, for $\isfresh{y}{\nu x.A, M, N}$, we have $A\sub{x}{y} \vDash M = N$,
and also
\item
$\theta \cpar P \vDash M = N$ whenever $M\theta \mathrel{=_E} N\theta$.
\end{itemize}
\end{definition}
The above ensures that the private names in an extended process
do not appear directly in $M$ or $N$, leaving only the possibility of 
using aliases in the domain of the active substitution in $M$ and $N$ to indirectly refer to 
private names. 
That is, $M$ and $N$ are recipes that must produce the same message, up to the equational theory $E$,
given the information recorded in the active substitution of the extended process.
As a simple example, we have $\mathopen{\nu x.}\left( {\sub{\prefix{0}\lambda}{x}}\circ{\sub{\prefix{1}\lambda}{h(x)}} \cpar P \right) \vDash h(\prefix{0}\lambda) = \prefix{1}\lambda$.

Now we can make a generic point about all reasonable notions of equivalence based on our structural operational semantics.
As explained in related work~\cite{Aubert2022e}, each alias has a location prefix, allowing each location to have its unique pool of aliases, thus ensuring that the choice of alias is localised and not impacted by choices of aliases made by concurrent threads.
For example, the following process has two transitions, labelled with $(\co{a}(\prefix{0}\lambda), \prefix{0}[])$ and $(\co{b}(\prefix{1}\lambda), \prefix{1}[])$ (\cf \autoref{def:independence} for a formal definition of those \emph{events}):
\begin{align*}
\mathopen{\nu x.}\left(
  {\sub{\prefix{0}\lambda}{x}} \cpar 0 \cpar {\cout{b}{h(x)}}
\right)
& \stld{\co{a}(\prefix{0}\lambda)}{ \prefix{0}[] }
\id \cpar \mathopen{\nu x.}\left({\cout{a}{x}} \cpar {\cout{b}{h(x)}}\right)
\dlts{\co{b}(\prefix{1}\lambda)}{ \prefix{1}[] }
\mathopen{\nu x.}\left(
  {\sub{\prefix{1}\lambda}{h(x)}} \cpar {\cout{a}{x}} \cpar 0
\right)
\shortintertext{Clearly, any reasonable semantics should equate the above process with the one below, where the only difference is that the parallel processes $\cout{a}{x}$ and $\cout{b}{h(x)}$
have been permuted (\eg exchanged their locations).}
\mathopen{\nu x.}\left(
  {\sub{\prefix{1}\lambda}{x}} \cpar {\cout{b}{h(x)}} \cpar 0 
\right)
& \stld{\co{a}(\prefix{1}\lambda)}{ \prefix{1}[] }
\id \cpar \mathopen{\nu x.}\left({\cout{b}{h(x)}} \cpar {\cout{a}{x}}\right)
\dlts{\co{b}(\prefix{0}\lambda)}{ \prefix{0}[] }
\mathopen{\nu x.}\left(
  {\sub{\prefix{0}\lambda}{h(x)}} \cpar 0 \cpar {\cout{a}{x}} 
\right)
\end{align*}
Notice that the events labelling the transitions differ only in the prefix string $\prefix{0}$ or $\prefix{1}$, but that this change impacts the domain of the active substitutions.
Therefore, when defining any notion of equivalence using this operational semantics, we must keep track of a substitution between aliases (which should be a bijection), thereby allowing for differences in prefixes and making the particular choice of alias irrelevant when performing equivalence checking.
%
%
\begin{definition}[alias substitution]
Alias substitutions $\rho$ extend to labels such that $(M\,N)\rho = M\rho\,N\rho$ and $(M(\alia))\rho = M\rho(\alia\rho)$, and $\tau\rho = \tau$. 
\end{definition}

The following function is just a convenience to 
pick out the domain of an active substitution.
This is useful since the domain remembers the set of aliases that have already been extruded.
\begin{definition}
We extend the domain function to 
extended processes such that
$\dom{\mathopen{\nu \vec{x}.}\left( \theta \cpar A \right)} = \dom{\theta}$.
\end{definition}

We make use of aliases substitution even for interleaving equivalences and similarities.
For example, the following\footnote{%
We \diff{color} what we want to stress or the \enquote{diff} with the previous definition or a definition indicated in footnote.%
} defines a notion of interleaving \enquote{presimilarity} (a term coined here to distinguish it from \enquote{similarity}, introduced in \autoref{def:i-sim}) that disregards the locations but requires the aliases to be substituted.  
\begin{definition}[interleaving presimilarity]
	\label{def:presim}
Let $\mathcal{R}$ be a relation between pairs of extended processes and $\rho$ be an alias substitution.
We say $\mathcal{R}$ is an i-presimulation whenever if $A \mathrel{\mathcal{R}^{\diff{\rho}}} B$, then:
\begin{itemize}
\item If $A \dlts{\pi}{u} A'$ then there exists \diff{$\rho'$}, $B'$, $u'$, $\pi'$ \st \diff{$\rho\mathclose{\restriction_{\dom{A}}} = \rho'\mathclose{\restriction_{\dom{A}}}$}, $B \dlts{\pi'}{u'} B'$, \diff{$\pi\rho' = \pi'$} and $A' \mathrel{\mathcal{R}^{\diff{\rho'}}} B'$.
\item If $A \vDash M = N$, then $B \vDash \diff{M\rho = N\rho}$.
\end{itemize}
We say process $P$ i-presimulates $Q$, and write $P \presimi{i} Q$, whenever
there exists a i-presimulation $\mathcal{R}$ such that
$\id \cpar P \mathrel{\mathcal{R}^{\id}} \id \cpar Q$.
\end{definition}
Notice that i-presimilarity $\presimi{i}$ is defined on processes: defining it on extended processes $A$ and $B$ require bijective alias substitutions $\rho$ such that $\dom{A}\rho = \dom{{B}}$ that complicate later definitions.

Now consider again the processes examined above
$\mathopen{\nu x.}\left({\cout{a}{x}} \cpar {\cout{b}{h(x)}}\right)$
and
$\mathopen{\nu x.}\left({\cout{b}{h(x)}} \cpar {\cout{a}{x}}\right)$.
They are mutually i-presimilar, \ie there exist two i-presimulations that relate them in each direction.
These presimulations involve building up a bijection on aliases
$\rho$ such that $\rho \colon \prefix{0}\lambda \mapsto \prefix{1}\lambda$
and
$\rho \colon \prefix{1}\lambda \mapsto \prefix{0}\lambda$.
By applying this bijection to the labels of each of the transitions
presented above, indeed the actions
of both processes,
$\co{a}(\prefix{0}\lambda)$ and $\co{a}(\prefix{1}\lambda)$ 
map to each other.
%
Observe also that the final states these processes reach
are 
$A = \mathopen{\nu x.}\left( {\sub{\prefix{0}\lambda}{x}}\circ{\sub{\prefix{1}\lambda}{h(x)}} \cpar 0 \cpar 0 \right)$
and
$B = \mathopen{\nu x.}\left( {\sub{\prefix{1}\lambda}{x}}\circ{\sub{\prefix{0}\lambda}{h(x)}} \cpar 0 \cpar 0 \right)$.
Since
$A \vDash h(\prefix{0}\lambda) = \prefix{1}\lambda$,
we also want this test to be satisfied by $B$, modulo the alias substitution $\rho$ that has been built by the presimilarity, \ie
$B \vDash (h(\prefix{0}\lambda))\rho = (\prefix{1}\lambda)\rho$, which indeed holds.
Notice that it is necessary to apply $\rho$ to the messages when checking that
equality tests are preserved, and that it must be applied before the active substitution.

One may ask whether it is possible to simply have a permutation of location prefixes, 
keeping alias variables the same.
Such an approach would not be sufficiently flexible to capture relations such as
\[\mathopen{\nu x.}\left({\cout{b}{h(x)}.\cout{a}{x}}\right) \presimi{i} \mathopen{\nu x.}\left({\cout{b}{h(x)}} \cpar {\cout{a}{x}}\right)
\qquad
\text{ and }
\qquad
\mathopen{\nu x.}\left({\cout{a}{x}} \cpar {\cout{x}{h(x)}}\right) \presimi{i} \mathopen{\nu x.}\left({\cout{a}{x}.\cout{x}{h(x)}}\right)
\text{.}
\]
In both examples, 
on one side there are two locations, and on the other there is only one location.
This helps explain why we employ a bijection between aliases and not only between locations. 

The above definition is an aesthetic preorder in that
we always match a positive test on the left with a positive test on right. 
The clause concerning equality tests effectively defines \enquote{static implication} 
proposed in related work on applied process calculi~\cite{Parrow2021}. 
However, there is a small gap compared to the standard simulation we expect for the $\pi$-calculus.
Indeed, the definition of presimilarity lets the following hold:
\[
\mathopen{\nu y.}\left( \cout{a}{x} + \cout{a}{y}\right) \presimi{i} \cout{a}{x}
\]
Therefore the above processes are mutually presimilar, since the other direction holds trivially.
The reason the above relation holds is that there is no equality that can distinguish the message $x$ from the private name $y$.
That is, both
\begin{align*}
\id \cpar \mathopen{\nu y.}\left(\cout{a}{x} + \cout{a}{y}\right) \dlts{\co{a}(\lambda)}{[0]} \mathopen{\nu y.}\left({\sub{\lambda}{x}} \cpar 0 \right) 
&& \text{ and }  &&
\id \cpar \mathopen{\nu y.}\left(\cout{a}{x} + \cout{a}{y}\right) \dlts{\co{a}(\lambda)}{[1]} \nu y.\left( {\sub{\lambda}{y}} \cpar 0 \right)
\end{align*}
can only be matched by
$\id \cpar \cout{a}{x} \dlts{\co{a}(\lambda)}{[]} {\sub{\lambda}{x}} \cpar 0$, 
and
there is no $M$ and $N$
such that 
$\nu y.\left( {\sub{\lambda}{y}} \cpar 0 \right) \vDash M = N$
and
 ${\sub{\lambda}{x}} \cpar 0 \nvDash M = N$.
Notice this is despite the fact that
 ${\sub{\lambda}{x}} \cpar 0 \vDash \lambda = x$,
but
$\nu y.\left( {\sub{\lambda}{y}} \cpar 0 \right) \nvDash \lambda = x$,
which would amount to  
$\nu y.\left( {\sub{\lambda}{y}} \cpar 0 \right)$
satisfying the inequality $\lambda \neq x$; hence such negative distinguishing tests
are not picked up on by presimilarity.


Intuitively, one can think of the above example modelling, with the left process, an \enquote{unreliable} channel (\ie output on channel $\co{a}$ can either be the intended message $x$ or anything else as $y$); whereas the right process is a reliable channel where the receiver would always get the intended message $x$.
Since we expect that in a conservative extension of the $\pi$-calculus
the above processes can be distinguished, we strengthen presimilarity to obtain \enquote{similarity}.
This strengthening amounts to demanding static equivalence, even when considering similarity preorders.
\begin{definition}[interleaving similarity]\label{def:i-sim}
Let $\mathcal{R}$ be a relation between pairs of extended processes and $\rho$ be an alias substitution.
We say $\mathcal{R}$ is a i-simulation whenever if $A \mathrel{\mathcal{R}^{\rho}} B$, then:
\begin{itemize}
\item If $A \dlts{\pi}{u} A'$ then there exists $\rho'$, $B'$, $u'$, $\pi'$ \st $\rho\mathclose{\restriction_{\dom{A}}} = \rho'\mathclose{\restriction_{\dom{A}}}$,
$B \dlts{\pi'}{u'} B'$, $\pi\rho' = \pi'$ and $A' \mathrel{\mathcal{R}^{\rho'}} B'$.
\item $A \vDash M = N$ \diff{iff} $B \vDash M\rho = N\rho$.
\end{itemize}
We say process $P$ i-simulates $Q$, and write $P \simi{i} Q$, whenever
there exists an i-simulation $\mathcal{R}$ such that $\id \cpar P \mathrel{\mathcal{R}^{\id}} \id \cpar Q$.
If in addition the relation is symmetric, \eg $A \mathrel{\mathcal{R}^{\rho}} B$ iff $B \mathrel{\mathcal{R}^{\rho^{-1}}} A$, then $P$ and $Q$ are i-bisimilar, written $P \sim_i Q$. 
\end{definition}
The notions of bisimilarity obtained from 
presimilarity and similarity concide, hence we see
similarity as presimilarity with a little of the power of
bisimilarity for equating tests.
Note that $\mathopen{\nu y.}\left(\cout{a}{x} + \cout{a}{y}\right)$ and $\cout{a}{x}$ are not i-similar, since there is a $\co{a}(\lambda)$-transition after which only the right side satisfies $\lambda = x$.

Definitions in related work on the applied $\pi$-calculus
do not require an alias substitution, as in the definition above.
Those papers~\cite{Abadi2018,Horne2021} allow the alias to be freely chosen, without indicating the location.
Notice also the location under the labelled transition is never used in these interleaving semantics.
The located aliases and location labels are however important for our non-interleaving equivalences, and for concurrency diamonds
required to extend techniques such as POR to the full applied $\pi$-calculus.

\section{
Using LATS to define semantics preserving duration or causality
}\label{sec:semantics-lats}

We now make the transition from interleaving to non-intereaving semantics.
The border between interleaving and non-interleaving semantics
was heavily debated in the early 1990's.
A common argument at the time was that
problems concerning non-interleaving semantics could be reduced to a problem
in terms of an interleaving semantics,
since processes such as $\nu x.\left({\cout{a}{x}} \cpar {\cout{a}{x}}\right)$ and 
$\nu x.\left({\cout{a}{x}.\cout{a}{x}}\right)$
could be distinguished by splitting each output actions into a \enquote{begin output}
and \enquote{end output} action and then considering the interleavings.
This view was eventually dispelled by van Glabbeek and Vaandrager~\cite{Glabbeek1997} (based on works, such as~\cite{aceto1994adding,GORRIERI1995272,Vogler1996}), who showed that, no matter how many times actions are split, one cannot obtain an interleaving semantics that preserves desirable properties of a non-interleaving semantics.

Their key example,
translated here to the $\pi$-calculus, is that there is an 
interleaving simulation relating the following processes.
\begin{equation}\label{eg:swap}
\mathopen{\nu c, d.}\left( \left( \cout{d}{d} \cpar \nu n.\cout{a}{n}.\cin{d}{z}.n(x)\right) \cpar \left( \cout{c}{c} \cpar \cin{c}{y}\right)\right)
\simi{i}
\mathopen{\nu c, d.}\left( \left( \cout{d}{d} \cpar \nu n.\cout{a}{n}.\cin{d}{z} \right) \cpar \left( \cout{c}{c} \cpar  \cin{c}{y}.n(x) \right)\right)
\end{equation}
Furthermore, even if we were to enhance similarity
with the power to split actions, these processes would still be related.
What is happening here is that 
when a $\tau$-transition both starts and terminates while
another $\tau$-transition is running,
the end of the longer and shorter $\tau$-transition can be swapped,
resulting in a behaviour that can be simulated on the right.
Such \enquote{swapping} semantics were investigated by Vogler~\cite{Vogler1996},
when investigating
the coarsest language theory robust
against splitting.

Although the above example preserves event splitting,
allowing it to hold can be considered problematic
since we confuse the beginning and end of two distinct events that happen
to be labelled in the same way. 
A notion of similarity allowing the above example to hold, neither preserves
the duration of events, nor the causal dependencies between events.
To see why, observe that the process on the left above 
has a $\tau$-transition that can start before any other event and terminate
after all events have finished, but there is no $\tau$-transition on the right
that can match that timing history.
In this section, we lift two truly non-interleaving semantics (ST and HP)
to the applied $\pi$-calculus that do preserve such properties.


\subsection{Independence and permutations of events}
\label{ssec:inde}

To define non-interleaving equivalences we make use of independence relations.
Structural independence, that looks only at the locations, is sufficient for calculi such as CCS.
However, for the $\pi$-calculus and its extensions, in addition,
so called \emph{link causality} should be  accounted for to determine whether an output must occur first before a subsequent event occurs.
\begin{definition}[independence]
	\label{def:independence}
	Define $\Locnoarg$  a function on location labels (\autoref{def:loclabel}) such that $\Loc{ \ell } = \left\{ \ell \right\}$ and $\Loc{\ell_0, \ell_1} = \left\{ \ell_0, \ell_1 \right\}$.
	The structural independence relation $\sIndy$ on location labels
	is the least relation defined by 
	$u_0 \sIndy u_1$
	whenever 
	for all locations $\ell_0 \in \Loc{u_0}$ and $\ell_1 \in \Loc{u_1}$, 
	there exist a string
	$\prefix{s} \in \left\{\prefix{0},\prefix{1}\right\}^*$
	and
	locations
	$\ell'_0, \ell'_1$,
	such that either:
	$\ell_0 = \prefix{s0}\ell'_0$
	and
	$\ell_1 = \prefix{s1}\ell'_1$;
	or 
	$\ell_0 = \prefix{s1}\ell'_0$
	and
	$\ell_1 = \prefix{s0}\ell'_1$.	
	Events $(\pi, u)$ are pairs of action labels $\pi$ and location labels $u$.
	The independence relation $\Indy$ on events
	is 
	the least symmetric relation such that
	$(\pi_0, u_0) \Indy (\pi_1, u_1)$
	whenever 
	$u_0 \sIndy u_1$
	and
	if $\pi_0 = \co{M}(\alia)$,
	then 
	$\isfresh{\alia}{\pi_1}$.
\end{definition}

Consider again \autoref{eg:swap}, where we present its executions as a graph where the events are nodes and edges represent dependencies (\ie the absence of independence).
Note $M$ is any message such that $\fa{M} \subseteq \left\{ \prefix{01}\lambda \right\}$, and results from an input.

\begin{tikzpicture}
	\node(orig) at (0, 0) {$(\co{a}(\prefix{01}\lambda), \prefix{01}[])$};
	\node[below= of orig] (1) {$(\tau , (\prefix{00}[], \prefix{01}[]))$};
	\node[below = of 1] (2){$(\prefix{01}\lambda\,M, \prefix{01}[])$};
	\node[right = of 1] (3){$(\tau, (\prefix{10}[], \prefix{11}[]))$};
	\draw[->] (orig) --(1);
	\draw[->] (1) --(2);
	\node[right = of 3] (vs){v.s.};
	\node [right = 6.5cm of orig](r1) {$(\co{a}(\prefix{01}\lambda), \prefix{01}[])$};
	\node [right = of r1](r2) {$(\tau, (\prefix{10}[], \prefix{11}[]))$};
	\node [below = 2.6cm of r1](r3) {$(\tau, (\prefix{00}[], \prefix{01}[]))$};
	\node [below = 2.6cm of r2](r4) {$(\prefix{01}\lambda\,M, \prefix{11}[])$};
	\draw[->] (r1) --(r3);	
	\draw[->] (r2) --(r4);	
	\draw[->] (r1) --(r4);	
\end{tikzpicture}

On the left above, observe that the rightmost $\tau$-transition  is independent from all other transitions,
while all other events in that diagram are dependent on each other.
In contrast, on the right above, both $\tau$-transitions are dependent on only one other event,
and independent of the others.
In what follows, we make precise what it means for 
the processes producing these events to be incomparable.

\subsection{ST-similarity and ST-bisimilarity, preserving duration} 
\label{ssec:st}

We define now ST semantics that preserve the duration of events, abstractly, without explicit time,
by providing mechanisms for modelling the start and termination of events.
To avoid confusion about which event
terminates at a particular moment, definitions of ST equivalences make use of a device to pair events that started at the same moment,
which is done by a relation over events in this work.
We define some simple auxiliary functions to work with relations and sets of events. 
\begin{definition}[auxiliary functions]
Given a relation over events $\ESS$,
we write $\dom{\ESS}$ and $\ran{\ESS}$ the sets of events forming the domain and range of $\ESS$, respectively.
Given an event $e$ and set of events $E$ we write $e \Indy E$ whenever for all $e' \in E$ we have $e \Indy e'$.
\end{definition}
Our definition of ST-similarity below enhances the definition of interleaving similarity 
such that we not only preserve the transitions, but also respect the fact that some events may have started already
and are running concurrently with the new event.
This is captured by ensuring that we only consider a transition labelled with event $(\pi, u)$ 
if the condition $(\pi, u) \Indy \dom{\ESS}$ holds, which ensures that all events currently running in $\ESS$
are independent of $(\pi, u)$.
We then demand that the corresponding transition, labelled with $(\pi', u')$,
is also independent of all events currently started, which is ensured by the condition $(\pi', u') \Indy \ran{\ESS}$.
Notice that the relation on events strongly associate $(\pi, u)$ and $(\pi', u')$,
and thus, when we appeal to the second clause below they will be removed from the relation simultaneously.\footnote{
Using a relation has the same effect as employing a bijection between the labels of events in other
formulations of ST-bisimilarity~\cite[p.~14]{Glabbeek90TRb}.%
}
This models the termination of the events.
Thus we only record in relation $\ESS$ those events that are concurrently running now,
which is suited to our independence relation that is only well-defined on transitions enabled in the same state or subsequent states.

\begin{definition}[ST-similarity]
	\label{def:st-sim}
Let $\mathcal{R}$ be a relation between pairs of extended processes, $\rho$ be an alias substitution, \diff{and $\ESS$ be a relation over events}.
We say $\mathcal{R}$ is an ST-simulation whenever if $A \mathrel{\mathcal{R}^{\rho, \diff{\ESS}}} B$, then:
\begin{itemize}
\item If $A \dlts{\pi}{u} A'$ 
and 
\diff{$(\pi, u) \Indy \dom{\ESS}$}
then there exists $\rho'$, $B'$, $u'$, and $\pi'$ \st
 $\rho\mathclose{\restriction_{\dom{A}}} = \rho'\mathclose{\restriction_{\dom{A}}}$,
$B \dlts{\pi'}{u'} B'$,
$\pi\rho' = \pi'$,
\diff{$(\pi', u') \Indy \ran{\ESS}$},
and
$A' \mathrel{\mathcal{R}^{\rho', \diff{\ESS \cup \left\{( (\pi, u) , (\pi', u'))\right\}}}} B'$.
\item \diff{If $\ESS' \subseteq \ESS$ then $A \mathrel{\mathcal{R}^{\rho, \ESS'}} B$}.
\item $A \vDash M = N$ iff $B \vDash M\rho = N\rho$.
\end{itemize}
We say process $P$ ST-simulates $Q$, and write $P \simi{ST} Q$, whenever
there exists a ST-simulation $\mathcal{R}$ \st 
$\id \cpar P \mathrel{\mathcal{R}^{\id, \emptyset}} \id \cpar Q$.
If in addition $\mathcal{R}$ is symmetric, 
\eg $A \mathrel{\mathcal{R}^{\rho, \ESS}} B$ iff $B \mathrel{\mathcal{R}^{\rho^{-1}, \ESS^{-1}}} A$, then $P$ and $Q$ are ST-bisimilar, written $P \bisimi{ST} Q$.
\end{definition}

Consider the following, which are i-bisimilar, but can be distinguished by ST-similarity.
\[
{\mathopen{\nu x.}\cout{a}{x}} \cpar {\mathopen{\nu x.}\cout{a}{x}}
  \nsimi{ST}
{\mathopen{\nu x.}\cout{a}{x} . \mathopen{\nu x.}\cout{a}{x}}
\]
To see why the above does not hold, 
observe that two events can be concurrently started on the left, but the second cannot be matched on the right.
That is, when playing the ST-simulation game, we reach the following states,
where $\rho \colon \prefix{0}\lambda \mapsto \lambda'$ and $\left(\co{a}(\prefix{0}\lambda), \prefix{0}[] \right) \ESS \left( \co{a}(\lambda'), []\right)$.
\[
\mathopen{\nu y.}\left( {\sub{\prefix{0}\lambda}{y}} \cpar 0 \cpar {\mathopen{\nu x.}\cout{a}{x}} \right)
  \mathrel{\mathrel{R}^{\rho, \ESS}}
\mathopen{\nu y.}\left( {\sub{\lambda'}{y}} \cpar  {\mathopen{\nu x.}\cout{a}{x}} \right)
\]
Now observe that the extended process on the left can perform an event $\left(\co{a}(\prefix{1}\lambda), \prefix{1}[] \right)$ independent of $\dom{\ESS}$, but the process on 
the right cannot perform any action independent of $\ran{\ESS}$.
From this we conclude that the above processes cannot be related by any ST-simulation.

We still however obtain many relations that also hold according to interleaving semantics.
For example, observe that the following holds.
\begin{align}
\nu x, y, z. (\cout{a}{x} . (\cout{b}{y} \cpar \cout{c}{z})) & \simi{ST} \nu x, y, z. (\cout{a}{x} . \cout{b}{y} \cpar \cout{c}{z})\text{.} \label{ex-st}
\shortintertext{Indeed, the left term's only transition}
\id \cpar \nu x, y, z. (\cout{a}{x} . (\cout{b}{y} \cpar \cout{c}{z})) & \dlts{\co{a}(\lambda)}{[]}  \nu x, y, z. ({\sub{\lambda}{x}} \cpar \cout{b}{y} \cpar \cout{c}{z}) \notag
\shortintertext{can easily be matched by the right  term}
\id \cpar \nu x, y, z. (\cout{a}{x} . \cout{b}{y} \cpar \cout{c}{z})  & \dlts{\co{a}(\prefix{0}\lambda)}{\prefix{0}[]}  \nu x, y, z. ({\sub{\prefix{0}\lambda}{x}} \cpar \cout{b}{y} \cpar \cout{c}{z}) \notag
\end{align}
and \(\rho : \lambda \mapsto \prefix{0}\lambda\), \(\ESS = \{((\co{a}(\lambda), []), (\co{a}(\prefix{0}\lambda), \prefix{0}[]))\}\) satisfies our definition.
Then, one needs to show that the resulting two terms are in \(\mathcal{R}^{\rho', \ESS}\) \emph{and} \(\mathcal{R}^{\rho', \emptyset}\).
For \(\mathcal{R}^{\rho', \ESS}\), since \(\nu x, y, z. ({\sub{\lambda}{x}} \cpar \cout{b}{y} \cpar \cout{c}{z})\)'s only transitions (with events $(\co{b}(\prefix{0}\lambda'), \prefix{0}[])$ and $(\co{c}(\prefix{1}\lambda'), \prefix{1}[])$) are \emph{not} independent with \(\dom{S} = (\co{a}(\lambda), [])\), they do not need to be matched by \(\nu x, y, z. ({\sub{\prefix{0}\lambda}{x}} \cpar \cout{b}{y} \cpar \cout{c}{z})\).
For \(\mathcal{R}^{\rho', \emptyset}\), it is straightforward to pair \( (\co{b}(\prefix{0}\lambda'), \prefix{0}[])\) and \( (\co{c}(\prefix{1}\lambda'), \prefix{1}[])\) with themselves, and to map \(\prefix{0}\lambda'\) and  \(\prefix{1}\lambda'\) to themselves.

Interestingly, two processes that are unrelated by ST-similarity can be in the limit identified even by ST-bisimilarity. 
Consider 
for example the following.
\[
{\mathopen{\nu x.}\cout{a}{x}} \cpar {\mathopen{\nu x.}\cout{a}{x}}
  \nsimi{ST}
\mathopen{\nu x.}\cout{a}{x} . \mathopen{\nu x.}\cout{a}{x}
\quad
\mbox{and yet}
\quad
\bang\mathopen{\nu x.}\cout{a}{x}
  \bisimi{ST}
{\bang(\mathopen{\nu x.}\cout{a}{x} . \mathopen{\nu x.}\cout{a}{x})}
\]
\label{relation}
To establish the equation on the right above, we construct the relation 
below
and prove that it is an ST-bisimulation by checking that each condition holds.
Firstly, $\ESS$ is downward closed, since it is not required to be defined for all $i \in \phi\cup\psi$.
When the right side leads, it can either start an action in a component that has not fired (in $L$ or greater than $n$),
or it can start a second component that is not blocked (\ie in $\phi$, such that $(\co{a}(\prefix{1^{i}0}\lambda), \prefix{1^{i}0}[]) \notin \ran{\ESS}$), either of which can 
be matched on the left by starting a new independent component.
When the left side leads it can only fire a new component, which can be matched by starting
a new component on the right. Those transitions are preserved by $\mathcal{R}^{\rho, \ESS}$;
notably, there can never be more concurrently started actions on the left than there are started components
on the right.
Let $\mathcal{R}$ be the least symmetric relation containing the following (upto $\equiv$).
\[
\mathopen{\nu \vec{z}.}\left(
\theta \cpar Q_{0} \cpar \ldots \left(Q_{m}
\cpar {\bang\mathopen{\nu x.}\cout{a}{x}}
\right) \ldots \right)
 \mathrel{\mathcal{R}^{\rho, \ESS}}
 \mathopen{\nu \vec{y}.}\left(
 \sigma \cpar P_{0} \cpar \ldots \left(P_{n}
 \cpar {\bang\mathopen{\nu x.}\cout{a}{x} . \mathopen{\nu x.}\cout{a}{x}}
 \right)\ldots \right)
\]
\begin{tabular}{l l l l}
$\left\{0, \ldots m\right\} = \chi \cup J$ & & 
$\left\{0, \ldots n\right\} = \psi \cup \phi \cup L$ \\
\multicolumn{2}{l}{with $\chi$ and $J$ disjoint and $m \notin J$} & 
\multicolumn{2}{l}{with $\psi$, $\phi$ and $L$ disjoint and $n \notin L$} \\
$Q_i = \begin{dcases*}
0  & if $i \in \chi$\\
\mathopen{\nu x.}\cout{a}{x}  & if $i \in J$
\end{dcases*}$
& & 
$P_i = \begin{dcases*}
	0  & if $i \in \psi$\\
	\mathopen{\nu x.}\cout{a}{x}   &  if $i \in \phi$\\
	\mathopen{\nu x.}\cout{a}{x}.\mathopen{\nu x.}\cout{a}{x}   & if $i \in L$
\end{dcases*}$
\\
 $\vec{z}_i =  \begin{dcases*}
z_i  & if $i \in \chi$\\
  \epsilon  & if $i \in J$
  \end{dcases*}$
&
$\vec{z} = \bigcup_{i = 0}^m \vec{z}_i$
& 
$\vec{y}_i =  \begin{dcases*}
	x_i, y_i  & if $i \in \psi$\\ 
	x_i  & if $i \in \phi$ \\
	\epsilon  & if $i \in L$
\end{dcases*}$
&
$\vec{y} = \bigcup_{i = 0}^n \vec{y}_i$
\\
$\theta_i = \begin{dcases*}
{\sub{\prefix{1^{i}0}\lambda}{z_i}} & if $i \in  \chi$\\
\id  &  if $i \in J$
\end{dcases*}$
&
$\theta = \prod_{i = 0}^m \theta_i$
&
$\sigma_i =  \begin{dcases*}
	{\sub{\prefix{1^{i}0}\lambda}{x_i}} \circ {\sub{\prefix{1^{i}0}\lambda'}{y_i}}  & if $i \in \psi$\\
	{\sub{\prefix{1^{i}0}\lambda}{x_i}} & if $i \in \phi$\\
	\id  &  if $i \in L$
\end{dcases*}$
&
$\sigma = \prod_{i = 0}^n \sigma_i$
\\
\multicolumn{4}{l}{with $ \rho : \dom{\theta} \rightarrow \dom{\sigma}$ any bijection such that
$(\prefix{1^{f(i)}0}\lambda)\rho = \begin{dcases*}
	 \prefix{1^{i}0}\lambda & if $i \in \phi$\\
	 \prefix{1^{i}0}\lambda' & if $i \in \psi$ 
\end{dcases*}$, for}\\
\multicolumn{4}{l}{
$f \colon \phi \cup \psi  \rightarrow \chi$ any injection
 and
$\begin{dcases*}
		(\co{a}( \prefix{1^{f(i)}0}\lambda ), \prefix{1^{f(i)}0}[]) \ESS (\co{a}(\prefix{1^{i}0}\lambda), \prefix{1^{i}0}[]) & only if $i \in \phi$\\
		(\co{a}( \prefix{1^{f(i)}0}\lambda), \prefix{1^{f(i)}0}[]) \ESS (\co{a}(\prefix{1^{i}0}\lambda'), \prefix{1^{i}0}[]) & only if $i \in \psi$
	\end{dcases*}$}
\end{tabular}

\subsection{History-Preserving similarity: preserving causality}
\label{ssec:hp}

Besides observing the duration of events as in ST semantics,
History-Preserving semantics observe also the partial order of causal dependencies between events.
We define here HP-similarity as a strengthening of our definition of ST-similarity
such that we observe not only independence but also dependence, thereby, step-by-step,
ensuring that exactly the same dependencies are satisfied by the events produced by both processes.
Technically this is achieved in the definition below, by partitioning the relation representing concurrently started
events $\ESS$ according to the firing event $(\pi, u)$ into: $\ESS_1$ consisting of events that are independent of the current event (\ie $(\pi, u) \Indy \dom{\ESS_1} $);
$\ESS_2$ consisting of those events that are not independent (\ie $(\pi, u) \notIndy \dom{\ESS_2}$).
Thus $\ESS_2$ is the minimal
set of events that must have terminated before the new event can proceed.
This partitioning must be reflected by the matching transition on the right, thereby preserving both independence and dependence.
Since only the independent events and the new event are retained at the next step, the relation over events always consists of independent events.
\begin{definition}[HP-similarity\footnote{This definition is "diffed" against \autoref{def:st-sim}. The clause \diff{\enquote{If $\ESS' \subseteq \ESS$ then $A \mathrel{\mathcal{R}^{\rho, \diff{\ESS'}}} B$.}} was replaced by the partitioning of events.}]
	\label{def:hp-sim}
Let $\mathcal{R}$ be a relation between pairs of extended processes, $\rho$ be an alias substitution, and $\ESS$ be a relation over events.
We say $\mathcal{R}$ is an HP-simulation whenever if $A \mathrel{\mathcal{R}^{\rho, \ESS}} B$, then:
\begin{itemize}
\item If $A \dlts{\pi}{u} A'$, \diff{$\ESS_1 \cup \ESS_2 = \ESS$}, 
\diff{$(\pi, u) \Indy \dom{\ESS_1}$} and 
\diff{$(\pi, u) \notIndy \dom{\ESS_2}$},
then there exists $\rho'$, $B'$, $u'$, and $\pi'$ \st
 $\rho\mathclose{\restriction_{\dom{A}}} = \rho'\mathclose{\restriction_{\dom{A}}}$,
$B \dlts{\pi'}{u'} B'$, 
$\pi\rho' = \pi'$,
\diff{$(\pi', u') \Indy \ran{\ESS_1}$},
$\diff{(\pi', u') \notIndy \ran{\ESS_2}}$,
and $A' \mathrel{\mathcal{R}^{\rho', \diff{\ESS_1} \cup \left\{( (\pi, u) , (\pi', u'))\right\}}} B'$.
\item $A \vDash M = N$ iff $B \vDash M\rho = N\rho$.
\end{itemize}
We say process $P$ is HP-simulated by $Q$, and write $P \simi{HP} Q$, whenever
there exists an HP-simulation $\mathcal{R}$ \st 
$\id \cpar P \mathrel{\mathcal{R}^{\id, \emptyset}} \id \cpar Q$.
If in addition $\mathcal{R}$ is symmetric,
then $P$ and $Q$ are HP-bisimilar, written $P \bisimi{HP} Q$.
\end{definition}

When we consider similarity the difference between ST-similarity and HP-similarity is clear.
For example, although \autoref{ex-st} proved the ST-similarity of the following, they are not HP-similar.
\[
\nu x, y, z. (\cout{a}{x} . (\cout{b}{y} \cpar \cout{c}{z})) \nsimi{HP} \nu x, y, z. (\cout{a}{x} . \cout{b}{y} \cpar \cout{c}{z})
\]
To see this, observe that when attempting to construct an HP-simulation we can reach the following pair of processes,
where $\rho \colon \lambda \mapsto \prefix{0}\lambda$ and
$(\co{a}(\lambda), []) \ESS (\co{a}(\prefix{0}\lambda), \prefix{0}[] )$.
\[
\mathopen{\nu x, y, z.}\left({\sub{\lambda}{x}} \cpar \cout{b}{y} \cpar \cout{c}{z} \right)
\mathrel{\mathcal{R}^{\rho, \ESS}}
\mathopen{\nu x, y, z.}\left( {\sub{\prefix{0}\lambda}{x}} \cpar \cout{b}{y} \cpar \cout{c}{z} \right)
\]
At this moment, the left side can perform a transition on channel $c$ that
is \emph{dependent} on $(\co{a}(\lambda), [])$ in $\dom{\ESS}$.
Yet, although the right  side can perform a transition on channel $c$,
it cannot match the dependency, since $(\co{c}(\prefix{1}\lambda), \prefix{1}[])$
and $(\co{a}(\prefix{0}\lambda), \prefix{0}[])$
are independent.

When we consider bisimilarity, the gap is more subtle for finite processes.
An example separating ST-bisimilarity from HP-bisimilarity 
is the following.
\begin{equation}\label{eg:confusion}
\mathopen{\nu a,b.}\left(
\left(
 \cout{a}{a}
\cpar 
 \left(
 \cin{a}{x}
 +
 \cin{b}{x}
 \right)
\right)
\cpar
 \cout{c}{c}.\cout{b}{b}
\right)
\bisimi{ST}
\mathopen{\nu a.}\left(
\left(
 \cout{a}{a}
\cpar 
 \cin{a}{x}
\right)
\cpar
 \cout{c}{c}
\right)
\end{equation}
To see that they are unrelated by HP-similarity (hence certainly unrelated by HP-bisimilarity),
observe that
the two processes can perform the following transitions
\begin{align*}
&& \id
\cpar
\mathopen{\nu a,b.}\left(
\left(
 \cout{a}{a}
\cpar \left(
 \cin{a}{x}
 +
 \cin{b}{x}
 \right)
\right)
\cpar
 \cout{c}{c}.\cout{b}{b}
\right)
& \dlts{\co{c}(\prefix{1}\lambda)}{\prefix{1}[]}
\mathopen{\nu a,b.}\left(
{\sub{\prefix{1}\lambda}{c}}
\cpar
\left(
 \cout{a}{a}
\cpar \left(
 \cin{a}{x}
 +
 \cin{b}{x}
 \right)
\right)
\cpar
 \cout{b}{b}
\right)\\
\text{and} && 
\id
\cpar
\mathopen{\nu a.}\left(
\left(
 \cout{a}{a}
\cpar 
 \cin{a}{x}
\right)
\cpar
 \cout{c}{c}
\right)
& \dlts{\co{c}(\prefix{1}\lambda)}{\prefix{1}[]}
\mathopen{\nu a.}\left(
{\sub{\prefix{1}\lambda}{c}}
\cpar
\left(
 \cout{a}{a}
\cpar 
 \cin{a}{x}
\right)
\cpar
 0
\right)\text{.}
\end{align*}
The relation on events at this moment is such that
$\left( \co{c}(\prefix{1}\lambda), \prefix{1}[] \right)
\mathrel{\ESS} 
\left( \co{c}(\prefix{1}\lambda), \prefix{1}[] \right)$
where alises are related by the identity function.
Notice now that 
$\mathopen{\nu a,b.}\left(
{\sub{\prefix{1}\lambda}{c}}
\cpar
\left(
 \cout{a}{a}
\cpar \left(
 \cin{a}{x}
 +
 \cin{b}{x}
 \right)
\right)
\cpar
 \cout{b}{b}
\right)$
can perform a transition labelled with
$(\tau, (\prefix{01}[1], \prefix{1}[]) )$,
which is not independent from
$\left( \co{c}(\prefix{1}\lambda), \prefix{1}[] \right)$;
yet, although the other process can perform a $\tau$-transition,
it cannot match the dependency constraints. 
In contrast, since ST-similarity would not require
dependency constraints to be matched, a matching $\tau$-transition
can be performed at the corresponding point in any ST-bisimulation game.

The distinction between ST and HP is less subtle when we consider replicated processes.
Consider
\[
\bang(\mathopen{\nu x.}\cout{a}{x} . \mathopen{\nu x.}\cout{a}{x}) 
\nsimi{HP}
\bang(\mathopen{\nu x.}\cout{a}{x})
\qquad
\text{ and yet }
\qquad
\bang(\mathopen{\nu x.}\cout{a}{x} . \mathopen{\nu x.}\cout{a}{x}) 
  \bisimi{ST}
\bang(\mathopen{\nu x.}\cout{a}{x})
\text{.}
\]

The latter relation above we have already established previously, p.~\pageref{relation}.
Now we attempt to construct an HP-simulation containing the relation on the left.
Observe that a possible first transition can be matched by both processes as follows.
\begin{align*}
\id \cpar \bang(\mathopen{\nu x.}\cout{a}{x} . \mathopen{\nu x.}\cout{a}{x})
& \dlts{\co{a}(\prefix{0}\lambda)}{\prefix{0}[]}
\nu y.\left( {\sub{\prefix{0}\lambda}{y}} \cpar
\mathopen{\nu x.}\cout{a}{x} \cpar
\bang(\mathopen{\nu x.}\cout{a}{x} . \mathopen{\nu x.}\cout{a}{x})
\right)
\\
\id \cpar \bang(\mathopen{\nu x.}\cout{a}{x})
& \dlts{\co{a}(\prefix{1^n0}\lambda)}{\prefix{1^n0}[]}
\nu y.\left( {\sub{\prefix{1^n0}\lambda}{y}} \cpar
0 \cpar
\left(\mathopen{\nu x.}{\cout{a}{x}}
\hdots
\left(
\mathopen{\nu x.}{\cout{a}{x}}
\cpar
\bang\mathopen{\nu x.}{\cout{a}{x}}
\right)\right)\right)
\end{align*}
At this point we have 
$( \co{a}(\prefix{0}\lambda), \prefix{0}[]) \ESS ( \co{a}(\prefix{1^n0}\lambda), \prefix{1^n0}[])$
and aliases substitution such that $\rho \colon \prefix{0}\lambda \mapsto \prefix{1^n0}\lambda$.
Then, 
$\nu y.\left( {\sub{\prefix{0}\lambda}{y}} \cpar
\mathopen{\nu x.}\cout{a}{x} \cpar
\bang(\mathopen{\nu x.}\cout{a}{x} . \mathopen{\nu x.}\cout{a}{x})
\right)$ can perform an event 
$( \co{a}(\prefix{0}\lambda'), \prefix{0}[])$
that is \emph{not independent} of 
$( \co{a}(\prefix{0}\lambda), \prefix{0}[])$, but 
the other process can only perform an independent transition, violating the condition of HP-similarity that the transition on the right must have the same dependencies. 


Similarly, we have 
$\bang \mathopen{\nu x, y.}\left( {\cout{a}{x}.\cout{b}{y}} + {\cout{b}{y}.\cout{a}{x}} \right)
\nbisimi{HP} 
\bang \nu x.{\cout{a}{x}}  \cpar \bang \nu x.{\cout{b}{x}}$
which are equated by the ST similarity.
We interpret these kinds of examples as follows.
From the perspective of the ST-semantics, executing the processes in an interleaved manner on one server that can be duplicated is the same as executing them on two servers that can be duplicated.
This is because the same duration of events can be achieved by both, and in some settings this may be the desirable effect.
However, this comes at the cost of a loss of awareness in the number of servers required (seen as resources), and of a sense of partition tolerance, since the right  process needs up to half as much servers as the left process requires to complete the same task.
This can be problematic if an attacker has the power to partition a system, \eg by DDoS on a connection link, thereby isolating a small number of servers from the rest.
In that situation, the difference picked out by HP-similarity becomes evident, and one can notice moreover that HP-similarity behaves the same in the finite case and in the limit.

There is related work on \enquote{causal} 
bisimilarity for the $\pi$-calculus~\cite{Boreale1998},
which is strictly 
finer than HP-bisimilarity.
This is because causal bisimilarity only accounts for structural causality
and not for link causality.
Thus, for example 
although
$
\mathopen{\nu n.}\left(
{\cout{a}{n}}
\cpar {\cin{n}{x}}
\right)
\bisimi{HP}
\mathopen{\nu n.}\left(
\cout{a}{n}.
{\cin{n}{x}}
\right)
$
holds, these processes are distinguished by causal bisimilarity, because \textcquote[p.~387]{Boreale1998}{there is both a subject and an object dependency between the actions \textins{in the former}, whereas in \textins{the latter} there is only an object dependency}.

\subsection{Discussion on ST and HP in the context of privacy}
\label{ssec:privacy}

We now revisit the essence of a privacy problem in the literature~\cite{Filimonov2019,Horne2021}.
The following compares two systems containing a process ready to respond to a message sent using a one-time key $k$,
\ie there is only one input action capable of responding to that key.
The left process allows processes in distinct locations to send a message using $k$, while on the right there is only one location with that capability.
Letting $P_{\pok} \triangleq \cin{b}{x}. \match{\snd{\dec{x}{k}} = \prhi}\cout{a}{ \enc{\pok}{k} }$, we have :
\begin{equation*}
\mathopen{\nu k.}\Big(
	(\nu r.\cout{a}{\enc{r, \prhi}{k}} \cpar
	\left(\nu m.{\cout{a}{m}}
	+ {\nu r.\cout{a}{\enc{r, \prhi}{k}}} \right)) \cpar
P_{\pok}~\Big)
\nsimi{i}
\mathopen{\nu k.}\Big(
	(
	\nu r.\cout{a}{\enc{r, \prhi}{k}}
	\cpar
	\nu m.\cout{a}{ m }
	) \cpar P_{\pok}
	~\Big)
\end{equation*}
The above processes are trace equivalent, yet these processes are distinguished by interleaving similarity
as indicated above.
Note that we assume a standard symmetic key Dolev-Yao equational theory $E$ such that $\dec{\enc{M}{K}}{K} \mathrel{=_E} M$,
$\fst{\pair{M}{N}} \mathrel{=_E} M$
and $\snd{\pair{M}{N}} \mathrel{=_E} N$.

Now compare this example above with the example below, where we essentially replicate some of the processes, and notice that, by doing so, these processes become i-bisimilar---they are even ST-bisimilar.
\begin{equation}
\label{eg:hp-priv}
\mathopen{\nu k.}\Big((
	\bang\nu r.\cout{a}{\enc{r, \prhi}{k}}
	\cpar
	\bang\nu m.\cout{a}{ m }
	) \cpar
P_{\pok}~\Big)
\tworels{\bisimi{ST}}{\nsimi{HP}}
\mathopen{\nu k.}\Big((
	\nu r.\cout{a}{\enc{r, \prhi}{k}}
	 \cpar
	 \bang\nu m.\cout{a}{ m }
	)
	 \cpar
	P_{\pok}~\Big)
\end{equation}

The problem is that there is no way for an observer to 
tell the difference between 
the output on channel $a$ after the match and a parallel
random output on channel $a$ (in the finite case all such parallel actions can be used up before performing the input, so it becomes clear whether or not $\enc{\pok}{k}$ is triggered, even without the attacker being able to read the message).
Of course, creating a channel for each process can be a solution to this
modelling problem~\cite{Horne2021}.
But the question we ask here is different: 
\emph{is the difference in locations picked up only by non-interleaving semantics?}.
%

The fact that the processes in \autoref{eg:hp-priv} are ST-bisimilar
shows that observing differences in the
duration of events does not affect the problem.
Indeed, while the output $\enc{\pok}{k}$ can only occur after the 
input, there is always another parallel action indistinguishable to the
attacker ready to fire for the same duration. 
Therefore ST-bisimilarity is not distinguishing sufficiently the localities
for this problem.

In contrast to the above, HP-similarity can detect the difference in localities.
This is because $\enc{\pok}{k}$ is triggered after the input,
and HP-similarity ensures that the same dependencies are preserved on
the right hand side of the simulation.

This problem is encapsulated by the following ST-bisimilar, but not mutually HP-similar, processes:
\[
 {\bang\nu n.\cout{a}{ n }} \cpar
{\cin{b}{x}.\nu n.\cout{a}{ h(n) }}
\tworels{\bisimi{ST}}{\nsimi{HP}}
 {\bang\nu n.\cout{a}{ n }} \cpar
{\cin{b}{x}}
\]

Hence, HP semantics is better at preserving structure, since we know that 
there is a success message (represented by $\enc{\pok}{k}$ here) caused by the input action, while ST semantics confuses this with other indistinguishable messages on channel $a$.

\subsection{Failure semantics}
\label{ssec:failure}

Considering simulations, not only bisimulation,
allows to explore more of the linear-time/branching-time spectrum.
For example, we can define ST failure similarity~\cite{Aceto1991}, which
extends ST-similarity such that if an action is 
enabled by the process on the right, 
then it should be enabled on the left.
\begin{definition}[STf-similarity\footnote{This definition is "diffed" against \autoref{def:st-sim}.}]
	\label{def:STf-sim}
Let $\mathcal{R}$ be a relation between pairs of extended processes, $\rho$ be an alias substitution, and $\ESS$ be a relation over events.
We say $\mathcal{R}$ is an STf-simulation whenever if $A \mathrel{\mathcal{R}^{\rho, \ESS}} B$, then:
\begin{itemize}
\item If $A \dlts{\pi}{u} A'$ 
and $(\pi, u) \Indy \dom{\ESS}$
then there exists $\rho'$, $B'$, $u'$, and $\pi'$ \st
 $\rho\mathclose{\restriction_{\dom{A}}} = \rho'\mathclose{\restriction_{\dom{A}}}$,
$B \dlts{\pi'}{u'} B'$, 
$\pi\rho' = \pi'$,
$(\pi', u') \Indy \ran{\ESS}$,
and $A' \mathrel{\mathcal{R}^{\rho', \ESS \cup \left\{( (\pi, u) , (\pi', u'))\right\}}} B'$.
\item
\diff{
If $B \dlts{\pi'}{u'} B'$ 
and $(\pi', u') \Indy \ran{\ESS}$
then there exists $\rho'$, $A'$, $u$ and $\pi$ \st
 $\rho\mathclose{\restriction_{\dom{A}}} = \rho'\mathclose{\restriction_{\dom{A}}}$,
$A \dlts{\pi}{u} A'$, 
$\pi\rho' = \pi'$,
and $(\pi, u) \Indy \dom{\ESS}$.
}
\item If $\ESS' \subseteq \ESS$ then $A \mathrel{\mathcal{R}^{\rho, \ESS'}} B$.
\item $A \vDash M = N$ iff $B \vDash M\rho = N\rho$.
\end{itemize}
We say process $P$ is STf-simulated by $Q$, 
and write $P \fsimi{ST} Q$,
 whenever
there exists an STf-simulation $\mathcal{R}$ such that
$\id \cpar P \mathrel{\mathcal{R}^{\id, \emptyset}} \id \cpar Q$.
\end{definition}
Tantalisingly,
 the above definition appears to preserve more dependencies than ST-similarity.
Not only can we detect differences in the branching structure, as expected for interleaving failure
similarity, but we can also detect the differences in the independence structure.
For instance we have:
\[
\nu x, y. (\cout{a}{x} . \cout{a}{y})  \nfsimi{ST} \nu x, y. (\cout{a}{x} \cpar \cout{a}{y})
\]

The distinguishing strategy is as follows.
Both processes are free to perform the first output on $a$
to reach the following indexed pair.
\[
\rho \colon \lambda \mapsto \prefix{0}\lambda
\quad
(\co{a}(\lambda), []) \ESS (\co{a}(\prefix{0}\lambda), \prefix{0}[]), \quad
\nu x, y. ({\sub{\lambda}{x}} \cpar \cout{a}{y}) \mathrel{\mathcal{R}^{\id, \ESS}}  \nu x, y. ({\sub{\prefix{0}\lambda}{x}} \cpar 0 \cpar \cout{a}{y})
\]
At this moment, the right hand side can perform a transition labelled with
event \((\co{a}(\prefix{1}\lambda), \prefix{1}[])\), since that event is independent of 
$(\co{a}(\prefix{0}\lambda), \prefix{0}[])$;
yet the process on the left cannot match this event.
Stated otherwise, the process on the left fails to perform
the next output on $a$ while the other output on $a$ is still
being performed, but the process on the right can.
This represents a failure measurable by
observing the concurrency of events.
Also,
$\nu x, y, z. (\cout{a}{x} . (\cout{b}{y} \cpar \cout{c}{z})) 
 \nfsimi{ST} \nu x, y, z. (\cout{a}{x} . \cout{b}{y} \cpar \cout{c}{z})$
since an action on channel $c$ is not enabled on the left initially.

Observing failures however does not allow
us to distinguish the processes in \autoref{eg:confusion}
nor in \autoref{eg:hp-priv},
since they are ST-bisimilar, hence 
mutually STf-similar.

We now adapt our privacy-inspired example of \autoref{ssec:privacy} to show the power of failure similarity.
The following are mutually ST-similar (and failure interleaving trace equivalent, which we do not define here), yet they are distinguished by STf-similarity.
Letting $P_{\perr} \triangleq  \cin{b}{x}. \match{\snd{\dec{x}{k}} \neq \prhi}\cout{a}{ \enc{\perr}{k} }$:
\begin{equation*}
	\mathopen{\nu k.}\Big((
		\nu r.\cout{a}{\enc{r, \prhi}{k}} \cpar
		\left(\nu m.{\cout{a}{s}}
		+ {\nu r.\cout{a}{\enc{r, \prhi}{k}}} \right)) \cpar
P_{\perr}~\Big)
\tworels{\simi{ST}}{\nfsimi{ST}}
	\mathopen{\nu k.}\Big((
		\nu r.\cout{a}{\enc{r, \prhi}{k}}
		\cpar
		\nu m.\cout{a}{ m }
		) \cpar P_{\perr}~\Big)
\end{equation*}

The difference compared to the example of \autoref{ssec:privacy} is that 
we can detect whether the outputs from the two locations are the same 
by \emph{not seeing an error} ($\perr$) after the input.
This kind of negative testing is part of the vocabulary
of failure semantics.
However, similarly to \autoref{eg:hp-priv}, if we include replication
then the processes become
ST-bisimilar, and hence cannot be 
distinguished by STf-similarity.
\begin{equation}
	\label{eg:HPf-priv}
	\mathopen{\nu k.}\Big((
		\bang\nu r.\cout{a}{\enc{r, \prhi}{k}}
		\cpar
		\bang\nu m.\cout{a}{ m }
		) \cpar P_{\perr}~\Big)
\tworels{\bisimi{ST}}{\nfsimi{HP}}
	\mathopen{\nu k.}\Big((
		\nu r.\cout{a}{\enc{r, \prhi}{k}}
		\cpar
		\bang\nu m.\cout{a}{ m }
		) \cpar
P_{\perr}~\Big)
\end{equation}
Despite the above processes being mutually STf-similar,
they are distinguished using HPf-similarity: 

\begin{definition}[HPf-similarity\footnote{This definition is "diffed" against \autoref{def:hp-sim}.}]
	\label{def:HPf-sim}
Let $\mathcal{R}$ be a relation between pairs of extended processes, $\rho$ be an alias substitution,
and $\ESS$ be a relation over events.
We say $\mathcal{R}$ is an HPf-simulation whenever if $A \mathrel{\mathcal{R}^{\rho, \ESS}} B$, then:
\begin{itemize}
\item If $A \dlts{\pi}{u} A'$,
$\ESS_1 \cup \ESS_2 = \ESS$,
$(\pi, u) \Indy \dom{\ESS_1}$ and 
$(\pi, u) \notIndy \dom{\ESS_2}$
then there exists $\rho'$, $B'$, $u'$, and $\pi'$ \st 
 $\rho\mathclose{\restriction_{\dom{A}}} = \rho'\mathclose{\restriction_{\dom{A}}}$
$B \dlts{\pi'}{u'} B'$, 
$\pi\rho' = \pi'$,
$(\pi', u') \Indy \ran{\ESS_1}$,
${(\pi', u') \notIndy \ran{\ESS_2}}$,
and $A' \mathrel{\mathcal{R}^{\rho', \ESS_1 \cup \left\{( (\pi, u) , (\pi', u'))\right\}}} B'$.
\item
\diff{If $B \dlts{\pi'}{u'} B'$,
$\ESS_1 \cup \ESS_2 = \ESS$,
$(\pi', u') \Indy \ran{\ESS_1}$
and $(\pi', u') \notIndy \ran{\ESS_2}$
then there exists $\rho'$, $A'$, $u$ and $\pi$ \st 
$\rho\mathclose{\restriction_{\dom{A}}} = \rho'\mathclose{\restriction_{\dom{A}}}$
$A \dlts{\pi}{u} A'$, 
$\pi\rho' = \pi'$,
$(\pi, u) \Indy \dom{\ESS_1}$,
and $(\pi, u) \notIndy \dom{\ESS_2}$.
}

\item $A \vDash M = N$ iff $B \vDash M\rho = N\rho$.
\end{itemize}
We say process $P$ is HPf-simulated by $Q$, and write $P \fsimi{HP} Q$, whenever
there exists an HPf-simulation $\mathcal{R}$ such that
$\id \cpar P \mathrel{\mathcal{R}^{\id, \emptyset}} \id \cpar Q$.
\end{definition}

To see why HPf-similarity can be used to distinguish the processes in \autoref{eg:HPf-priv},
observe that after inputing a message encrypted with $k$
in two possible ways,
we can tell that, 
on the right,
in at least one case
there will be an output message on channel $a$ \emph{that is
dependent on the input}.
Yet on the left it is possible, in both cases, that
neither can perform such an output.
An important part of this is the dependencies of the error message
that we do not see, since all messages are indistinguishable to 
the attacker who does not know $k$, and hence cannot tell by looking
at the message whether it is an error message.

Interestingly, anything coarser than HPf-similarity would not distinguish the processes
in \autoref{eg:HPf-priv}, since we use branching-time (so they are pomset failure trace equivalent\footnote{
We do not define failure trace semantics in this paper, 
however it is easy to see how to obtain it via our approach to located aliases in \autoref{sec:interleaving}
combined with classic definitions~\cite{Aceto1991,Vogler1991}.
}), failures (so they are HP-similar), \emph{and} causality preservation (so they are ST-bisimilar): we need all the features of HPf-similarity.

\section{Comparison to located bisimulations}
\label{sec:located-bisim}
This section compares our definitions to located equivalences, to help explain some less obvious design decisions.
Early work on LATS for CCS defined a notion of bisimilarity preserving independence~\cite{Mukund1992}.
A key difference compared to our definition of HP-bisimilarity is that 
all events are accumulated in a history of events, whereas our definition
remembers only those events that are currently active, and need not yet have terminated.
Remembering all events may appear to simplify things,
but we explain in this section that doing so gives rise to located equivalences 
that preserve the location of events, but forget about causal dependencies.
To see this, consider the following processes, which are equivalent, even with respect to HP-bisimilarity.
\[
L_1
\triangleq
\nu b.\left(
\cout{a}{a}.\cout{b}{b}
\cpar 
\cin{b}{x}.\cout{c}{c}
\right)
\qquad
\bisimi{HP}
\qquad
\nu b.\left(
\cout{b}{b}
\cpar 
\cout{a}{a}.\cin{b}{x}.\cout{c}{c}
\right)
\triangleq
L_2
\]
To see why these processes are 
HP-bisimilar observe there are only three possible transitions for both processes,
and one choice of alias substitution, as follows.
\[
\id \cpar L_1 \dlts{\cout{a}{\prefix{0}\lambda}}{\prefix{0}[]}
\dlts{\tau}{(\prefix{0}[], \prefix{1}[])}
\dlts{\cout{c}{\prefix{1}\lambda}}{\prefix{1}[]}
\qquad
\id \cpar  L_2
\dlts{\cout{a}{\prefix{1}\lambda}}{\prefix{1}[]}
\dlts{\tau}{(\prefix{0}[], \prefix{1}[])}
\dlts{\cout{c}{\prefix{1}\lambda'}}{\prefix{1}[]}
\qquad
\rho: \prefix{0}\lambda \mapsto \prefix{1}\lambda
\qquad
\rho: \prefix{1}\lambda \mapsto \prefix{1}\lambda'
\]
There are no other transitions (modulo renaming $\lambda$, of course), and none of these events can be permuted.
Notice that after each step the next transition is not independent of the currently started transitions, hence any started event must be removed from the set of active independent transitions $\ESS_1$ for the 
game to continue.
Therefore, we can pair the four states of these processes to form an HP-bisimulation.

In contrast, for the established located bisimilarities based on a LATS,
the set of all events that have happened is accumulated in $\ESSTEE$, and the independence of our LATS is preserved over all events.
That is, we remember all pairs of events, and preserve independence everywhere, as captured by the following definition.
\begin{definition}[$I$-consistent relation]
For some symmetric relation over events $I$, 
an $I$-consistent relation over a set of events, say $\ESSTEE$, is such that
if 
$(e_0, d_0) \in \ESSTEE$
and
$(e_1, d_1) \in \ESSTEE$
then 
 $e_0 \mathrel{I} e_1$ iff $d_0 \mathrel{I} d_1$.
\end{definition}
The definition above can be instantiated with any notion of independence over events,
such as $\sIndy$ or $\Indy$ as in \autoref{def:independence}, denoted here by $I$.

Now if we accumulate all pairs of events for our example above we obtain, after three transitions,
the relation over events $\ESSTEE$ defined as follows.
\[
\left( \cout{a}{\prefix{0}\lambda}, \prefix{0}[] \right)
\ESSTEE
\left( \cout{a}{\prefix{1}\lambda}, \prefix{1}[] \right)
\quad
\left( \tau, (\prefix{0}[], \prefix{1}[]) \right)
\ESSTEE
\left( \tau, (\prefix{0}[], \prefix{1}[]) \right)
\quad
\left( \cout{c}{\prefix{1}\lambda}, \prefix{1}[] \right)
\ESSTEE
\left( \cout{c}{\prefix{1}\lambda'}, \prefix{1}[] \right)
\]
Taking the relation $I$ to be $\Indy$, we have that
the above is not $\Indy$-consistent, since 
$\left( \cout{a}{\prefix{0}\lambda}, \prefix{0}[] \right) \sIndy
\left( \cout{c}{\prefix{1}\lambda}, \prefix{1}[] \right)$ holds
but
$\left( \cout{a}{\prefix{1}\lambda}, \prefix{1}[] \right)
\sIndy 
\left( \cout{c}{\prefix{1}\lambda'}, \prefix{1}[] \right)$
does not. 

An immediate consequence of the above is that 
the definition of bisimulation based on $I$-consistency, defined below, 
preserves the location of events more strongly than $HP$-bisimilarity, which preserves causal relationships.
Indeed when we take $I$ to be $\sIndy$, obtaining $\sIndy$-bisimilarity,
we obtain a located bisimilarity and located bisimilarities and HP-bisimilarities are known to be incomparable. 

\begin{definition}[$I$-similarity]
	\label{def:l-sim}
Let $\mathcal{R}$ be a relation between pairs of extended processes and $\rho$ be an alias substitution.
We say $\mathcal{R}$ is an $I$-simulation whenever if $A \mathrel{\mathcal{R}^{\rho, \ESSTEE}} B$, then:
\begin{itemize}
\item $\ESSTEE$ is $I$-consistent.

\item If $A \dlts{\pi}{u} A'$ 
then there exists $\rho'$, $B'$, $u'$, and $\pi'$ \st
$\rho\mathclose{\restriction_{\dom{A}}} = \rho'\mathclose{\restriction_{\dom{A}}}$,
$B \dlts{\pi'}{u'} B'$, 
$\pi\rho' = \pi'$,
and $A' \mathrel{\mathcal{R}^{\rho', \ESSTEE \cup \left\{( (\pi, u) , (\pi', u'))\right\}}} B'$.
         
\item $A \vDash M = N$ iff $B \vDash M\rho = N\rho$.
\end{itemize}
We say process $P$ $I$-simulates $Q$, and write $P \simi{I} Q$, whenever
there exists an I-simulation $\mathcal{R}$ \st 
$\id \cpar P \mathrel{\mathcal{R}^{\id, \emptyset}} \id \cpar Q$.
If in addition $\mathcal{R}$ is symmetric, then $P$ and $Q$ are $I$-bisimilar, written $P \bisimi{I} Q$.
\end{definition}

In a sense, it is just a coincidence that for CCS,
the above definition exploits nicely the independence relation of CCS, which coincides with $\sIndy$ since there is no link causality, and hence is strongly linked to the definition of a LATS for CCS. 
If we try to use $\Indy$-bisimilarity, using the full independence relation $\Indy$ from \autoref{def:independence}, that accounts for link causality, we end up with an awkward relation.
This has to do with the fact that independence for a LATS for the $\pi$-calculus must respect link causality, which means, for example, that the following processes are $\Indy$-bisimilar:
\[
\mathopen{\nu n.}\left( {\cout{a}{n}} \cpar n(x) \right)
\quad
\bisimi{\Indy}
\quad
\nu n. \cout{a}{n}. n(x) 
\]
This is because 
for both processes, the two events can only execute in one order, and neither is independent of the other, hence the set of events are $\Indy$-consistent.
Yet these processes are not $\sIndy$-bisimilar, since their pairing $\ESS$ is not $\sIndy$-consistent.
This is rather troubling when juxtapositioned with the observation that the following are not $\Indy$-bisimilar.
\[
\mathopen{\nu n.}\left( {\cout{a}{n}} \cpar n(x).\cout{\pok}{\pok} \right)
\quad
\nbisimi{\Indy}
\quad
\nu n. \cout{a}{n}. n(x).\cout{\pok}{\pok}
\]
Similarly to the above we have that the three events may only be fired in
a given order.
However, the resulting relation over events is not $\Indy$-consistent, since the first and
third events are independent for the left process above,
but are not independent for the right process above.
This seems strange that the first event of the
sub-process $n(x).\cout{\pok}{\pok}$ is somehow not location-sensitive,
yet the second is. 
To us, this is morally  broken, hence $\bisimi{\Indy}$ is ill-defined.
On the other hand $\sim_{\sIndy}$ consistently distinguishes these
two examples, where the former involves two locations while the latter involves only one location.

Indeed $\sim_{\sIndy}$ is the notion of bisimilarity
that would be obtained from the notion of trace equivalence implemented in the equivalence checking
tool DeepSec~\cite{Cheval2019}.
They call their equivalence session equivalence
and define it for a fragment of the applied $\pi$-calculus only.
It is clear that a notion of trace equivalence
that ensures that the events in compared traces are $\sIndy$-consistent
is the session equivalence of DeepSec.
Intuitively, this is because session equivalence forms a bijection
between processes in distinct locations and matches the behaviours
in each location, which is exactly what $\sIndy$-consistency would demand.
Interestingly, that tool employs partial order reduction to 
improve equivalence checking; which is evidence that POR might be 
lifted to other notions of equivalence defined in this paper.

Thus, for the $\pi$-calculus and its extensions, there seems to be no real connection between
$\Indy$ and located bisimilarity; effectively we throw away part of the LATS to obtain a located bisimilarity~\cite{Sangiorgi1996}.
The above observations help explain two things. 
Firstly, why we chose to target equivalences related to ST-similarity and HP-similarity rather than located bisimilarities in this work.
Secondly, why our definitions are more complicated than those for located bisimilarities for CCS in the literature.

\section{Conclusion}
\label{sec:conclusion}

Having introduced a LATS for the applied $\pi$-calculus~\cite{Aubert2022e},
we have shown that a world of non-interleaving operational semantics opens up for value passing process calculi.
Notably, by using the independence relation (\autoref{def:independence})
of a LATS, we capture ST-bisimilarity (\autoref{def:st-sim}) and HP-bisimilarity (\autoref{def:hp-sim}) that reflect
correctly link causality, which were not preserved by established located
bisimilarities for the $\pi$-calculus.
Both semantics have their merits:
for infinite processes, ST-semantics are very close to interleaving semantics,
while being naturally compatible with the independence relation of a LATS;
while HP-semantics better preserves the testing of finite subcomponents, even when we consider
limits and infinite process.
\autoref{eg:hp-priv} showed that HP-similarity is able to detect attacks that are
detectable using interleaving similarity for finite systems, yet
are not detectable even by the strictly more powerful  
ST-bisimilarity when we take limits.
This observation is reinforced 
in~\autoref{eg:HPf-priv}
where we show that HP failure similarity
picks up on attacks that would be missed
by anything coarser in any dimension (ST-bimilarity, HP-similarity, or even pomset failure traces).
Since HP-bisimilarity would equally pick up on the attacks,
we suggest HP-bisimilarity may be a good choice for security.

Having these definitions opens up formal and practical questions.
It is non-trivial to verify that these definitions are the same
as what we would expect if we pass via the more denotational world
of event structures, configuration structures,
or ST-structures~\cite{Glabbeek2009,Johansen2016}.
It is also non-trivial to provide characterisations using tests and modal logics~\cite{Hennessy95}. 
What is fairly clear is that the relationship between these notions,
since we start with the minimal notion of presimilarity
and grow from there, providing separating examples at each step.
The practical questions are more pressing, in particular, whether
we can make use of ST- and HP-semantics in tools for protocol verification.

\paragraph{Acknowledgements}
The definitions in this paper are introduced to support an invited talk by the second author
at EXPRESS/SOS on proving privacy properties using bisimilarity.
We thank the organisers 
Valentina Castiglioni
and
Claudio Antares Mezzina
for this invitation.



\bibliographystyle{eptcs}
\bibliography{bib/bib}

\appendix

\end{document}